\documentclass{PoS}

\usepackage{epsfig}

\PoS{PoS(LAT2006)157}

\title{Testing UV-filtered (``fat-link'') clover fermions}

\ShortTitle{Testing UV-filtered clover fermions}

\author{Stefano Capitani\\
        Universit\"at Graz, Institut f\"ur Physik, A-8010 Graz, Austria\\
        E-mail: \email{stefano.capitani\,(AT)\,uni-graz.at}}

\author{\speaker{Stephan D\"urr}\\
        Universit\"at Bern, ITP, Sidlerstrasse 5, CH-3012 Bern, Switzerland\\
        E-mail: \email{durr\,(AT)\,itp.unibe.ch}}

\author{Christian Hoelbling\\
        Bergische Universit\"at Wuppertal, Gaussstr.\,20, D-42119 Wuppertal,
        Germany\\
        E-mail: \email{christian.holbling\,(AT)\,cern.ch}}

\abstract{We investigate filtered clover fermions, built from fat gauge links,
both in one-loop perturbation theory and in numerical simulations. We use a
variety of filtering recipes (APE, HYP, EXP, HEX), some of which are suitable
for a HMC with dynamical fermions. A generic filtering together with a
(fat-link) clover term yields fermions with much reduced chiral symmetry
breaking.}

\FullConference{XXIVth International Symposium on Lattice Field Theory\\
                July 23-28, 2006\\
                Tucson, Arizona, USA}

\newcommand{\al}{\alpha}
\newcommand{\be}{\beta}
\newcommand{\ga}{\gamma}
\newcommand{\de}{\delta}

\newcommand{\ka}{\kappa}

\newcommand{\si}{\sigma}

\newcommand{\pa}{\partial}
\newcommand{\ovr}{\over}
\newcommand{\til}{\tilde}

\renewcommand{\dag}{^\dagger}
\newcommand{\<}{\langle}
\renewcommand{\>}{\rangle}

\newcommand{\bdm}{\begin{displaymath}}
\newcommand{\edm}{\end{displaymath}}
\newcommand{\bea}{\begin{eqnarray}}
\newcommand{\eea}{\end{eqnarray}}
\newcommand{\beq}{\begin{equation}}
\newcommand{\eeq}{\end{equation}}

\newcommand{\mr}{\mathrm}

\newcommand{\MeV}{\,\mr{MeV}}
\newcommand{\GeV}{\,\mr{GeV}}

\newcommand{\fm}{\,\mr{fm}}

\def\twocolcc#1{\multicolumn{2}{|c|}{#1}}

\begin{document}


\section{Overview}


Standard Wilson fermions are fairly fast to simulate, since the pertinent
Dirac operator
\beq
D_\mr{W}(x,y)={1\ovr2}\sum_\mu
\Big\{
(\ga_\mu-I) U_\mu(x)\de_{x+\hat\mu,y}-
(\ga_\mu+I) U_\mu\dag(x-\hat\mu)\de_{x-\hat\mu,y}
\Big\}
+{1\ovr2\ka}\de_{x,y}
\label{def_wils}
\eeq
is sparse, and they have a strong point by preserving flavor.
The main disadvantage is that they break chiral symmetry.
There are two established procedures that can ameliorate the latter
\begin{itemize}
\itemsep-2pt
\item
``$O(a)$-improvement'': $D_\mr{W}  \to  D_\mr{SW}=
D_\mr{W}-{c_\mr{SW}\ovr2}\sum_{\mu<\nu}\si_{\mu\nu}F_{\mu\nu}\;\de_{x,y}$
\item
``UV-filtering'' or ``link-fattening'': $U_\mu(x) \to  U_\mu^\mr{APE,HYP,...}$
and $F_{\mu\nu}(x) \to  F_{\mu\nu}^\mr{APE,HYP,...}$
\end{itemize}
and the purpose of this talk is to highlight the fact that by \emph{combining}
both approaches, one can considerably reduce the amount of chiral symmetry
breaking.
It is easy to obtain residual quark masses of the order of $30\MeV$ to
$100\MeV$ without any need for tuning.

In \cite{Capitani:2006ni} we extend and systematize earlier studies
\cite{DeGrand:1998jq,Bernard:1999kc,Stephenson:1999ns,Zanotti:2001yb,
DeGrand:2002va,DeGrand:2002vu}.
In particular, Ref.\,\cite{Bernard:1999kc} has proven useful to us, since it
is the very first account on fat-link perturbation theory.


\section{UV-filtering (``smearing'', ``link-fattening'') recipes}


The idea of any UV-filtered fermion action is that one would carry on a
smoothed copy of the actual gauge field and evaluate the Dirac operator on that
background.
This yields a new fermion action which differs from the old one by
terms which are simultaneously \emph{ultralocal} and \emph{irrelevant}.
The term ``UV-filtered'' indicates that such an action is less sensitive to
the UV fluctuations of the gauge background.
One may also speak of ``fat-link'' actions, but one should avoid the word
``improved'', since the Symanzik class with typically $O(a^2)$ cut-off effects
is maintained.

Obviously, there is a large amount of freedom.
One needs to decide on the smoothing recipe (APE, HYP, etc.), on the parameter
($\al^\mr{APE},\al_{1,2,3}^\mr{HYP}$) and on the number of iterations,
$n^\mr{iter}$.
With $D_\mr{SW}$ one may either build just the clover term from smoothed links
(FLIC fermions \cite{Zanotti:2001yb}), or use the same type of smoothing in the
covariant derivative, too (as we do).
In any case, with fixed $(\al,n^\mr{iter})$ the filtered (``fat-link'') action
is in the \emph{same universality class} as the usual (``thin-link'') version.

We compare a total of 16 actions.
This comes, since we start from the Wilson ($c_\mr{SW}\!=\!0$) and the
Sheikholeslami-Wohlert (``clover'', $c_\mr{SW}\!=\!1$) actions.
Then we investigate four recipes, APE, HYP, EXP, HEX.
The first two are well known, the third is the ``stout'' recipe of
Ref.\,\cite{Morningstar:2003gk}, and the fourth one is the hypercubic nesting
idea with this EXP/stout inside -- see \cite{Capitani:2006ni} for details.
Finally, we choose $n^\mr{iter}=1,3$.
To prevent further proliferation, we stay with one parameter per recipe
\beq
\begin{array}{lcl}
\mr{APE\;with\;\;} \al^\mr{APE}=0.6 &
\stackrel{\mr{PT}}{\longleftrightarrow} &
\mr{EXP\;with\;\;} \al^\mr{EXP}=0.1 \;\mbox{[``stout'']}
\\
\mr{HYP\;with\;\;} \al^\mr{HYP}=(0.75,0.6,0.3) &
\stackrel{\mr{PT}}{\longleftrightarrow} &
\mr{HEX\;with\;\;} \al^\mr{HEX}=(0.125,0.15,0.15)
\end{array}
\eeq
thereby exploiting a one-to-one relationship (in 1-loop PT)
APE$\,\leftrightarrow\,$EXP and ditto for HYP$\,\leftrightarrow\,$HEX.

In fact, filtered (``fat-link'') PT is not much harder than naive
(``thin-link'') PT. For APE \cite{Bernard:1999kc}
\beq
A_\mu^{(n)}(q)=\sum_\nu\,
\Big(
[1-{\al\ovr2(d\!-\!1)}\hat q^2]^n\,
(\de_{\mu,\nu}-{\hat q_\mu \hat q_\nu\ovr \hat q^2})
+{\hat q_\mu \hat q_\nu\ovr \hat q^2}
\Big)
\,A_\nu(q)
\eeq
and this means that in $d$ dimensions the transverse part of the gauge field
gets multiplied with a form-factor
$f_\mr{APE}^{(n)}(\hat q^2)=[1-{\al\ovr2(d\!-\!1)}\hat q^2]^n$
for $n$ iterations \cite{Bernard:1999kc}, where
$\hat{q}_\mu={2\ovr a}\sin({a\ovr2}q_\mu)$.


\section{Critical mass in 1-loop PT}


\begin{table}
\begin{center}
\begin{tabular}{|c|ccccc|}
\hline
{               }&   thin\,link&  1\,APE  &  2\,APE  &  3\,APE  &    1\,HYP  \\
\hline
$c_\mr{SW}\!=\!0$&{\bf51.43471}& 13.55850 &  7.18428 &  4.81189 &{\bf6.97653}\\
$c_\mr{SW}\!=\!1$&{\bf31.98644}&  4.90876 &  1.66435 &  0.77096 &{\bf1.98381}\\
$c_\mr{SW}\!=\!2$&     1.10790 & -7.11767 & -5.48627 & -4.23049 &   -4.41059 \\
\hline
\end{tabular}
\end{center}
\vspace{-4mm}
\caption{Additive mass shift $S$ in 1-loop PT for standard Wilson or clover
fermions on Wilson glue and after APE/EXP or HYP/HEX filtering. The uncertainty
is of order one in the last digit quoted.}
\end{table}

One way to assess the amount of chiral symmetry breaking is to consider the
magnitude of the additive mass renormalization.
In PT it has the expansion [with $C_F=4/3$ for $SU(3)$]
\beq
am_\mr{crit}=\Sigma_0=-{g_0^2\ovr16\pi^2}C_F S+O(g_0^4) \qquad [<0]
\eeq
with $S$ given in Tab.\,1.
Focusing on the numbers in bold print, one sees that standard clover
improvement diminishes $S$ by a factor 1.6.
On the other hand, one HYP step reduces it by a factor 7.4.
The interesting news is that by combining both strategies one obtains a
factor 26, that is more than the product of the two.
Hence we reach the conclusion that (at least in 1-loop PT) the two ingredients
clover-improvement and link-fattening pile up to suppress $-am_\mr{crit}$
quite drastically.


\section{Renormalization factors in 1-loop PT}


\begin{table}
\begin{center}
\begin{tabular}{|c|ccc|}
\hline
$c_\mr{SW}\!=\!0$&  thin\,link&  1\,APE  &  1\,HYP  \\
\hline
$z_S$            &   12.95241 &  1.12593 & -1.78317 \\
$z_P$            &   22.59544 &  5.28288 &  0.51727 \\
$z_V$            &   20.61780 &  6.39810 &  3.38076 \\
$z_A$            &   15.79628 &  4.31963 &  2.23054 \\
\hline
$(z_P\!-\!z_S)/2$&{\bf4.82152}&  2.07848 &{\bf1.15022}\\
$z_V\!-\!z_A$    &{\bf4.82152}&  2.07847 &{\bf1.15022}\\
\hline
\end{tabular}
\begin{tabular}{|c|ccc|}
\hline
$c_\mr{SW}\!=\!1$&  thin\,link&  1\,APE  &  1\,HYP  \\
\hline
$z_S$            &   19.30995 &  4.11106 & -0.03678 \\
$z_P$            &   22.38259 &  4.80364 &  0.12845 \\
$z_V$            &   15.32907 &  3.31243 &  1.38517 \\
$z_A$            &   13.79274 &  2.96614 &  1.30255 \\
\hline
$(z_P\!-\!z_S)/2$&{\bf1.53632}&  0.34629 &{\bf0.08262}\\
$z_V\!-\!z_A$    &{\bf1.53633}&  0.34629 &{\bf0.08262}\\
\hline
\end{tabular}
\caption{Coefficient $z_X$ in formula (4.2) for the renormalization factor
$Z_X$ with $X=S,P,V,A$ after 1 step of APE or HYP filtering. Entries are for
$c_\mr{SW}\!=\!0$ Wilson fermions (left) or $c_\mr{SW}\!=\!1$ clover fermions
(right).}
\end{center}
\end{table}

Another way to assess the amount of chiral symmetry breaking is to consider
\beq
\begin{array}{c}
\<.|O_j^\mr{cont}(\mu)|.\>=\sum_k Z_{jk}(a\mu) \<.|O_k^\mr{latt}(a)|.\>
\\[2mm]
Z_{jk}(a\mu)
=\de_{jk}-{g_0^2\ovr16\pi^2}(\Delta_{jk}^\mr{latt}-\Delta_{jk}^\mr{cont})
=\de_{jk}-{g_0^2\ovr16\pi^2}C_F z_{jk}
\end{array}
\eeq
for the point-like scalar/pseudoscalar densities and vector/axialvector
currents.
Then
\beq
\begin{array}{rcl}
Z_S(a\mu)=1-{g_0^2\ovr4\pi^2}\Big[{z_S\ovr3}-\log(a^2\mu^2)\Big]+...
&\quad\quad&
Z_V=1-{g_0^2\ovr12\pi^2}z_V+...
\\[2mm]
Z_P(a\mu)=1-{g_0^2\ovr4\pi^2}\Big[{z_P\ovr3}-\log(a^2\mu^2)\Big]+...
&\quad\quad&
Z_A=1-{g_0^2\ovr12\pi^2}z_A+...
\end{array}
\label{def_zX}
\eeq
with $z_S,z_P,z_V,z_A$ given in Tab.\,2.
A first check whether $(z_P\!-\!z_S)/2$ equals $z_V\!-\!z_A$ is successful.

Furthermore, in view of the overlap action satisfying
$z_S^\mr{ov}\!=\!z_P^\mr{ov},z_V^\mr{ov}\!=\!z_A^\mr{ov}$, it is clear that
this (common) figure [in the last two lines of Tab.\,2] quantifies the amount
of chiral symmetry breaking.
Focusing on the numbers in bold print, one sees that standard clover
improvement diminishes $z_V\!-\!z_A$ by a factor 3.14.
On the other hand, one HYP step reduces it by a factor 4.19.
The interesting news is that by combining both strategies one obtains a
factor 58.4, that is more than the product of the two.
Hence we reach the conclusion that (at least in 1-loop PT) the two ingredients
clover-improvement and link-fattening pile up to suppress
${1\ovr2}(z_P\!-\!z_S)=z_V\!-\!z_A$ quite drastically.

\begin{figure}
\epsfig{file=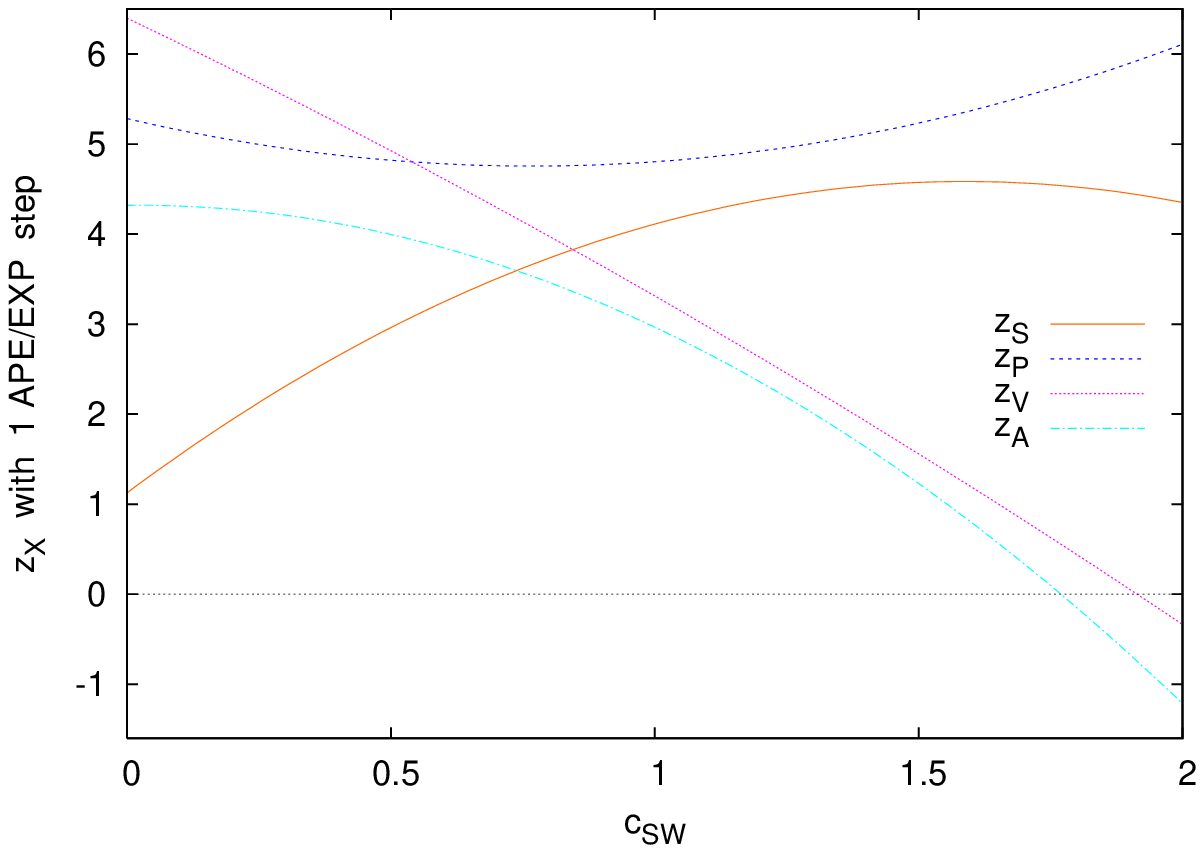,width=7.5cm}
\epsfig{file=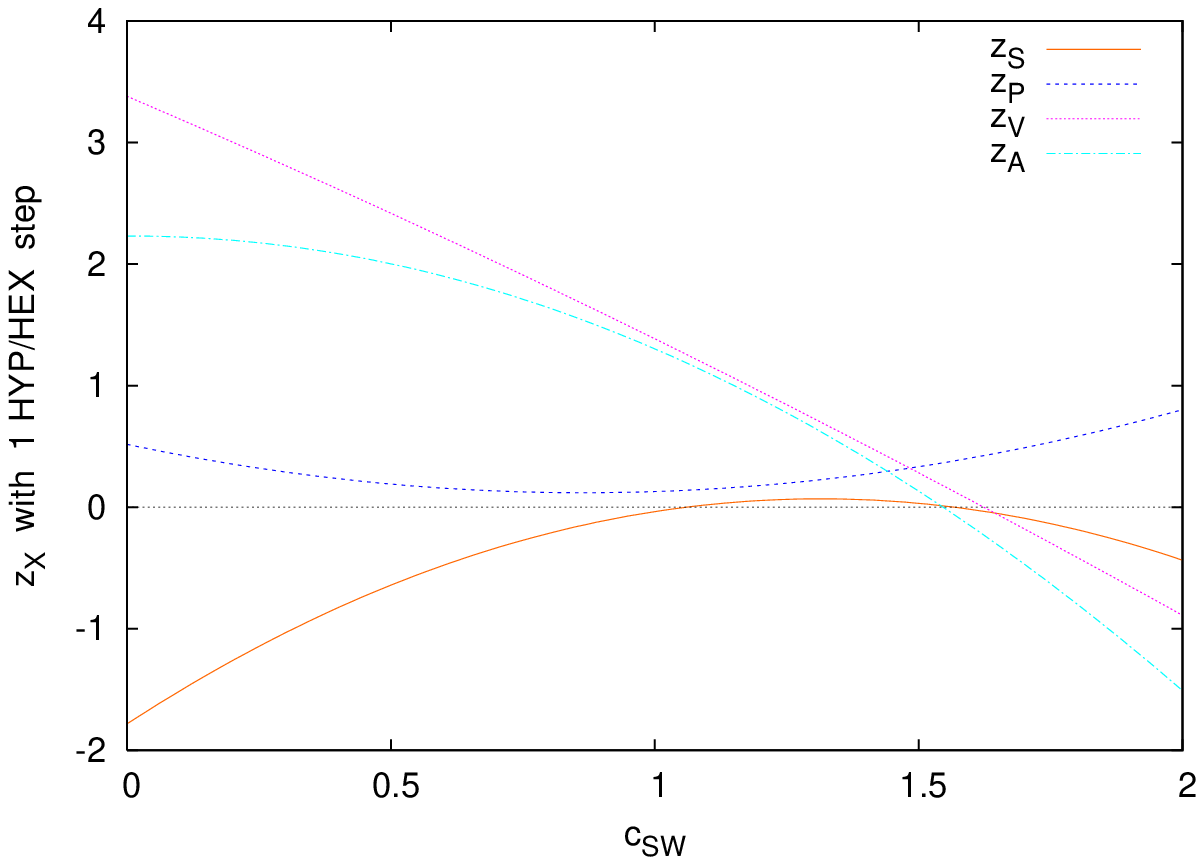,width=7.5cm}
\caption{Finite pieces $z_{S,P,V,A}$ of the $Z_X$ for 1\,APE and 1\,HYP
fermions as a function of $c_\mr{SW}$.}
\end{figure}

Finally, one may consider how the $z_X$ (with $X=S,P,V,A$) depend on
$c_\mr{SW}$.
In 1-loop PT one finds quadratic polynomials with details given in Fig.\,1.
For the action with one APE step $c_\mr{SW}\!=\!1.2648$ is the point where
$z_S$ is closest to $z_P$ (and hence $z_V$ closest to $z_A$).
For the action with one HYP step the minimal amount of chiral symmetry breaking
is realized through $c_\mr{SW}\!=\!1.1653$.
This gives some hope that a one-loop improved action
($c_\mr{SW}\!=\!1+\mr{const}\,g_0^2$) or even a tree-level improved action
($c_\mr{SW}\!=\!1$) with some vigorous filtering might have decent chiral
properties.


\section{Irrelevance of tadpole resummation}


\begin{table}
\begin{center}
\begin{tabular}{|c|cccccccc|}
\hline
  & $0.12$ & $0.24$ & $0.36$ & $0.48$ & $0.6$  & $0.72$ & $0.84$ & $0.96$ \\
\hline
1 & 6.99558& 5.19536& 3.77414& 2.73191& 2.06867& 1.78443& 1.87918& 2.35292\\ 
2 & 5.44459& 3.26311& 2.05185& 1.39240& 1.02644& 0.85574& 0.94215& 1.50761\\
3 & 4.32095& 2.22832& 1.31922& 0.89113& 0.66450& 0.55028& 0.63677& 1.39614\\
4 & 3.49281& 1.62650& 0.94513& 0.64620& 0.48918& 0.40469& 0.49903& 1.64011\\
5 & 2.87228& 1.25138& 0.72799& 0.50519& 0.38709& 0.32019& 0.42898& 2.29060\\
\hline
\end{tabular}
\caption{Tadpole diagram in Landau gauge [value to be multiplied with
$g_0^2C_F/(16\pi^2)$] in 1-loop PT for various $(n^\mr{iter},\al^\mr{APE})$.
The corresponding ``thin-link'' value is 9.174788.}
\end{center}
\end{table}

For ``thin-link'' actions it is customary to split $S$ in Landau gauge into
two contributions
\beq
S/(16\pi^2)=-\Sigma_0/(g_0^2C_F)=
-[\mr{sunset }]_0/(g_0^2C_F)
-[\mr{tadpole}]_0/(g_0^2C_F)
\eeq
where the sunset piece carries (in 1-loop PT) all dependence on $c_\mr{SW}$
and is (at least for $0\!\leq\!c_\mr{SW}\!\leq\!2$) ``small'', while the
tadpole piece is ``large''.
Hence it makes sense to resum the latter \cite{Lepage:1992xa}.

With filtering the situation is different, as a glimpse at Tab.\,3 reveals.
For the APE/EXP recipes (the columns simultaneously refer to
$\al^\mr{APE}\!=\!0.12,0.24,...$ and $\al^\mr{EXP}\!=\!0.02,0.04,...$) any
intermediate choice of $(\al,n^\mr{iter})$ renders the tadpole contribution
much smaller than in the unfiltered case (where it is 9.174788).
With the tadpole and hence $S$ being small for a generic filtering, there is no
need to resum the tadpole contributions.
In summary, the irrelevance of tadpole resummation for ``fat-link'' actions
gives us hope that PT in $g_0^2$ might converge nicely for UV-filtered actions.


\section{Non-perturbative tests}


\begin{table}
\begin{center}
\begin{tabular}{|c|ccccc|}
\hline
$\be$            & 5.846 & 6.000 & 6.136 &  6.260 & 6.373 \\
$L/a$            &   12  &   16  &   20  &    24  &   28  \\
$a^{-1}\,[\GeV]$ & 1.590 & 2.118 & 2.646 &  3.177 & 3.709 \\
$n_\mr{conf}$    &   64  &   32  &   16  &     8  &    4  \\
\hline
\end{tabular}
\caption{Matched ($\be$, $L/a$) combinations; $n_\mr{conf}$ is the number of
configurations per filtering and mass.}
\end{center}
\end{table}

Given the good chiral properties of UV-filtered actions in 1-loop PT, it is
useful to check to which extent they are realized non-perturbatively.
To this end we performed a quenched study with $a^{-1}$ ranging from $1.6\GeV$
to $3.7\GeV$ in a fixed physical volume $V\!=\!(1.5\fm)^4$.
We use the Wilson gauge action with the parameters of Tab.\,4.
The bare Wilson and PCAC masses are defined as
\bea
m^\mr{W}&=&
m_0-m_\mr{crit}  \quad\mbox{where}\quad
am_0={1\ovr2\ka}-4\;,\;am_\mr{crit}={1\ovr2\ka_\mr{crit}}-4
\\
m^\mr{PCAC}&=&
{\<\bar\pa_\mu[A_\mu^a(x)+ac_A\bar\pa_\mu P^a]O^a(0)\> \ovr 2\<P^a(x)O^a(0)\>}
\eea
with $\bar\pa$ the symmetric derivative, and the renormalized VWI and AWI quark
masses follow via
\bea
m^\mr{VWI}(\mu)&=&Z_m(a\mu)(1+b_m am^\mr{W})m^\mr{W}
\\
m^\mr{AWI}(\mu)&=&{Z_A\ovr Z_P(a\mu)}\,
{1+b_A am^\mr{W}\ovr 1+b_P am^\mr{W}}\,m^\mr{PCAC}
\;.
\eea
We use $c_\mr{SW}\!=\!0,1$ and the tree-level improvement coefficients
$b_X\!=\!1,c_{V,A}\!=\!0$.
It turns out that $b_m\!=\!-{1\ovr2}$ leads to unacceptable fits.
Since we cannot afford one more free parameter, we choose to set $b_m\!=\!0$.
From considering the $m^\mr{PCAC}$ versus $m_0$ we extract the inverse slope
$\til{Z}_A=Z_A\,Z_S/Z_P$ and the horizontal offset $-am_\mr{crit}$.
The idea is now to compare them to predictions from 1-loop PT.


\section{Rational fits for $-am_\mr{crit}$}


We know that asymptotically $-am_\mr{crit}\to{S\ovr12\pi^2}g_0^2$ with $S$
given in Tab.\,1.
Fitting our data with
\beq
-am_\mr{crit}={c_1g_0^2+c_2g_0^4\ovr1+c_3g_0^2}
\label{rat_mcrit}
\eeq
we have two options.
We may set $c_1$ to its perturbative value and adjust $c_2,c_3$.
Or we may fit all three coefficients and compare the fitted $c_1$ to its
perturbative prediction.
It turns out that the first option leads (at our couplings) to unacceptable
fits, hence we are left with the second one.
In Fig.\,2 we plot our non-perturbative values of $-am_\mr{crit}$ versus
$g_0^2$ for our 16 actions.
The results of the rational fits (\ref{rat_mcrit}) are included and one sees
that they give a decent description of the data.

\begin{figure}
\epsfig{file=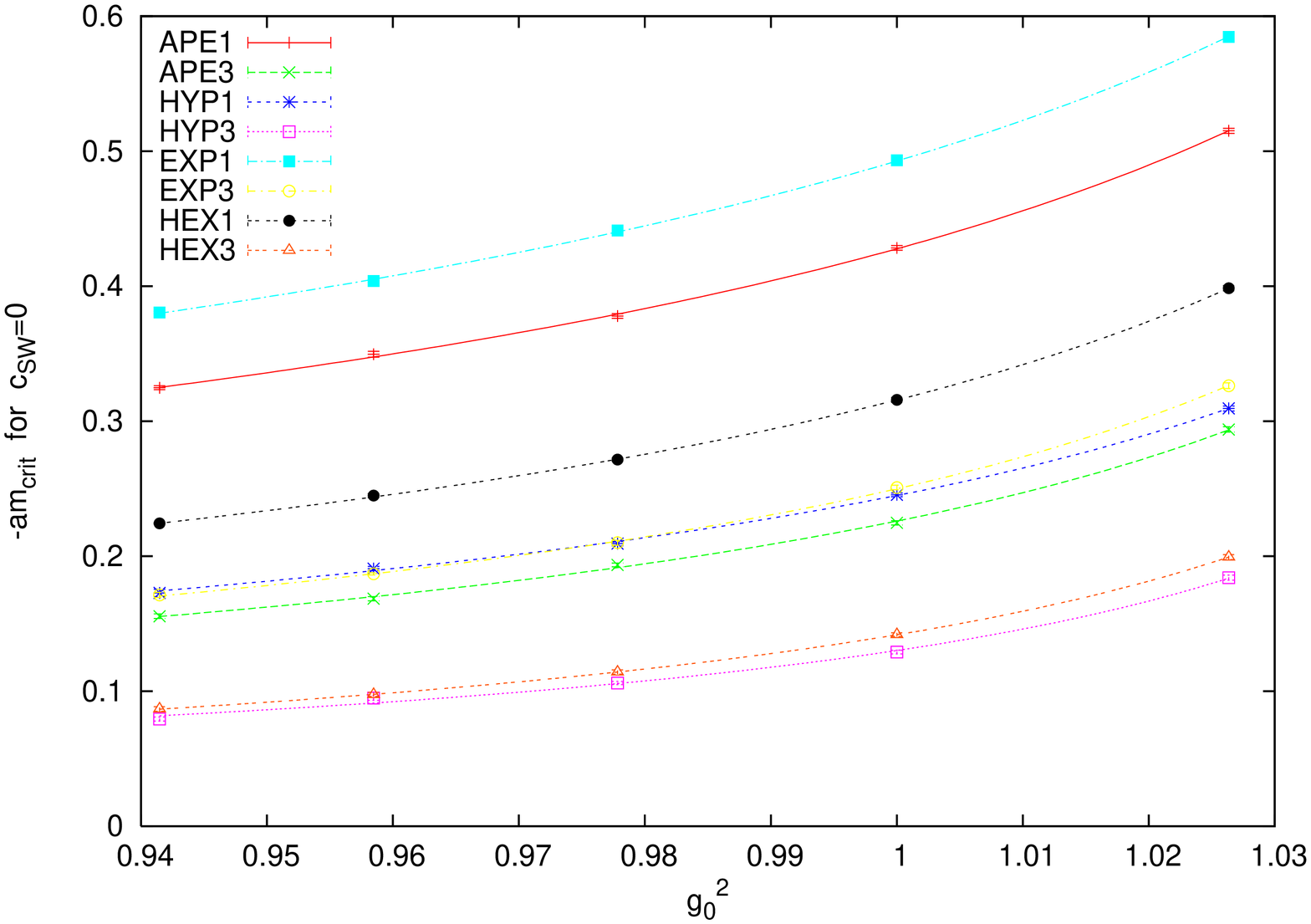,height=7.5cm,angle=-90}
\epsfig{file=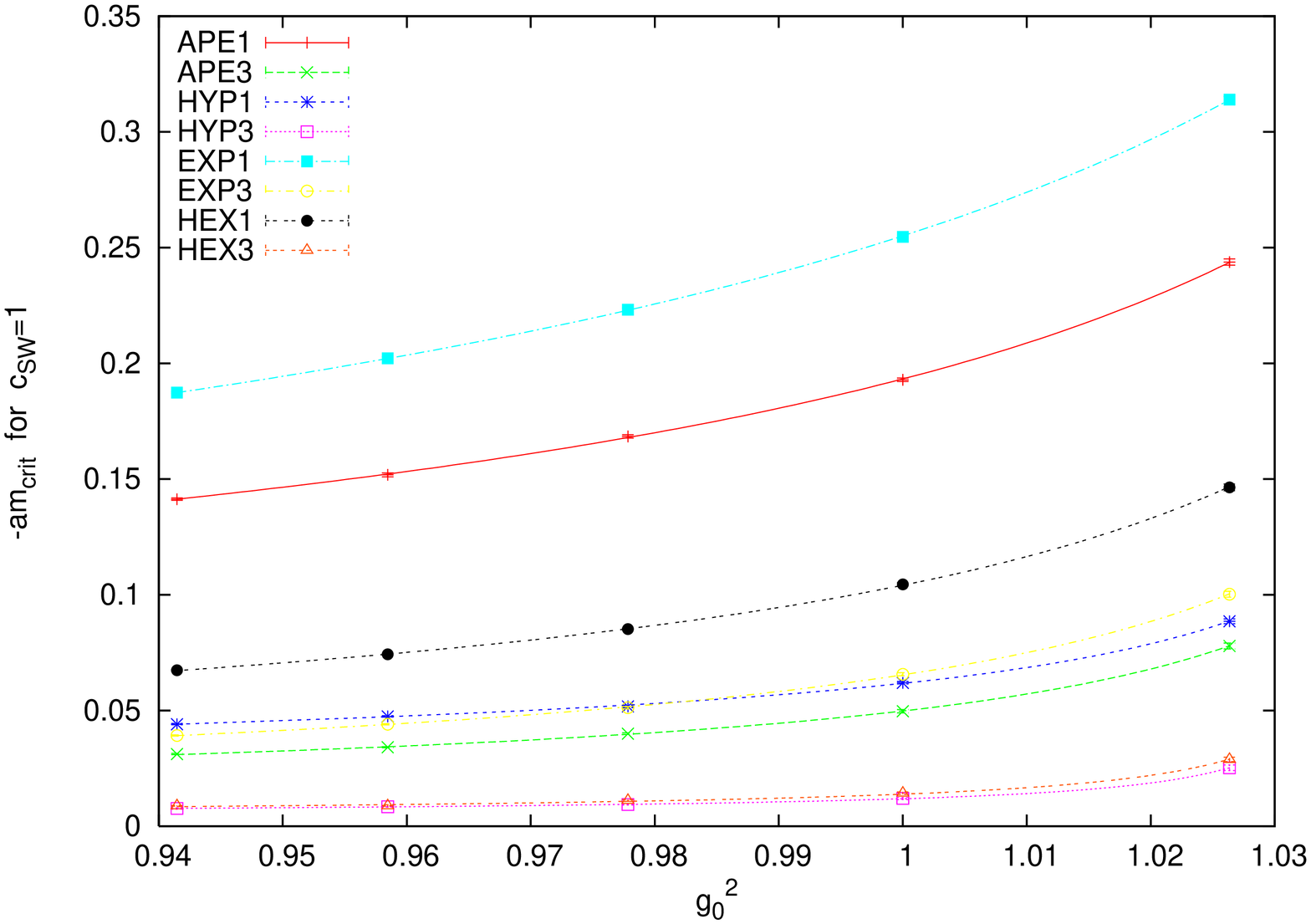,height=7.5cm,angle=-90}
\caption{$-am_\mr{crit}$ vs.\ $g_0^2$ for Wilson ($c_\mr{SW}\!=\!0$, left) and
clover ($c_\mr{SW}\!=\!1$, right) fermions with our 8 recipes.}
\end{figure}

\begin{table}
\begin{center}
\begin{tabular}{|cc|cc|cc|}
\hline
        &        & \twocolcc{$c_\mr{SW}=0$}& \twocolcc{$c_\mr{SW}=1$} \\
\hline
\twocolcc{pert.} &    \twocolcc{0.114480}  &    \twocolcc{0.0414467}  \\
 1\,APE & 1\,EXP &  0.213(12) & 0.252(12)  &  0.0909(28) & 0.1094(20) \\
\hline
\twocolcc{pert.} &    \twocolcc{0.040629}  &    \twocolcc{0.0065096}  \\
 3\,APE & 3\,EXP &  0.077(14) & 0.083(07)  &  0.0172(15) & 0.0171(09) \\
\hline
\twocolcc{pert.} &    \twocolcc{0.058906}  &    \twocolcc{0.0167502}  \\
 1\,HYP & 1\,HEX &  0.095(14) & 0.121(04)  &  0.0338(12) & 0.0332(16) \\
\hline
\twocolcc{pert.} &    \twocolcc{  ---   }  &    \twocolcc{  ---  }    \\
 3\,HYP & 3\,HEX &  0.034(15) & 0.026(01)  &  0.0060(02) & 0.0060(15) \\
\hline
\end{tabular}
\caption{Fitted $c_1$ values in (7.1) compared to the 1-loop prediction
$S/(12\pi^2)$ with the $S$ taken from Tab.\,1.}
\end{center}
\end{table}

The pertinent $c_1$ coefficients are collected in Tab.\,5,
along with the 1-loop prediction with $S$ taken from Tab.\,1.
Comparing the perturbative and the non-perturbative values, one may say that
they are close on an absolute scale (set by the unfiltered action), since all
$c_1$ are small.
However, on a relative scale, the two differ significantly --- typically, the
non-perturbative $c_1$ is larger than the perturbative one by a factor 2--3.
In spite of this disagreement, the non-perturbative data still show a
consistency $c_1^\mr{APE}\!\simeq\!c_1^\mr{EXP}$ and ditto for
$c_1^\mr{HYP}\!\simeq\!c_1^\mr{HEX}$, as predicted in PT.
We find this amusing, in particular in view of the fact that the pertinent
curves in Fig.\,2 are not close at all.
In summary, we would say that there are some encouraging signs, but there is
no quantitative agreement of $c_1$ with 1-loop PT in our range of couplings.

\begin{table}
\begin{center}
\begin{tabular}{|c|ccccc|}
\hline
{}& (5.846) & 6.000 & 6.136 &  6.260 & 6.373 \\
\hline
$m_\mr{res}^\mr{3\,APE}[\MeV]$&(144)&111&107&108&113\\
$m_\mr{res}^\mr{3\,HYP}[\MeV]$& (47)& 27& 25& 26& 27\\
\hline
\end{tabular}
\caption{Residual mass [defined via $m^\mr{PCAC}$] for our couplings. We
estimate that the error is of order $5\MeV$.}
\end{center}
\end{table}

We did similar fits for $\til{Z}_A$ and $am_\mr{res}$.
In the former case the filtering achieves $\til{Z}_A\!\simeq\!1$, but the
deviation from $1$ is not adequately described in 1-loop PT.
In the latter case, the situation is analogous to $-am_\mr{crit}$, which is no
surprise, since $\til{Z}_A\!\simeq\!1$ implies
$|m_\mr{crit}|\!\simeq\!m_\mr{res}$.
Our $am_\mr{res}$ [defined as the AWI mass at $m_0\!=\!0$] differs from the
version that is standard in the domain-wall community; nonetheless, it might
be interesting to compare.
We collect our results, converted into physical units, in Tab.\,6.
The first striking feature is that, after abandoning the coarsest lattice, they
are almost independent of the coupling.
The second surprise is that the tree-level improved 3\,HYP action achieves
$m_\mr{res}\!\simeq\!O(30\MeV)$.
Clearly, this lies well above of the residual masses that can be achieved with
domain-wall fermions.
On the other hand, it is much smaller than the residual mass of an unfiltered
Wilson or clover action, typically $O(1\GeV)$.
Our hope is that the small $m_\mr{res}$ of UV-filtered clover quarks leads
to good scaling properties and reduces the CPU time requirements to obtain a
predefined accuracy of phenomenologically interesting observables in the
continuum.


\section{Summary}


We close with highlighting some key properties of UV-filtered (``fat-link'')
clover actions:
\begin{enumerate}
\itemsep -1pt
\item
UV-filtering of $D_\mr{SW}$ yields a legal action for any fixed
$(\al^\mr{recipe},n^\mr{iter})$.
\item
1-loop PT suggests that the series in $g_0^2$ at $O(a)$ converges well,
without tadpole resummation.
\item
Maybe even the the non-perturbative $O(a)$ ambiguities in $c_V,c_A$ will be
gone.
\item
Chiral symmetry breaking is reduced: $m^\mr{res}\!\simeq\!30-100\MeV$
and ${1\ovr2}(z_P\!-\!z_S)=z_V\!-\!z_A\!\ll\!1$.
\item
With EXP/stout filtering (and maybe 1-loop $c_\mr{SW}$) $D_\mr{SW}$ is
ready for dynamical simulations.
\end{enumerate}



\end{document}